\documentclass[prl,twocolumn,showpacs,10pt]{revtex4-1}
\usepackage{textcomp}
\usepackage{amsmath}
\usepackage{amsfonts}
\usepackage{calc}
\usepackage{mathrsfs}
\usepackage{amssymb}
\usepackage{amsmath}
\usepackage{array}
\usepackage{color}
\usepackage{bm}
\usepackage{graphicx}
\usepackage{hyperref}

\newcommand{\e}{\epsilon}
\newcommand{\nhat}{\hat{n}}
\newcommand{\E}{\mathcal{E}}
\newcommand{\B}{\mathcal{B}}

\begin{document}
\title{Second-order gravitational self-force} 
\author{Adam Pound$^1$} 
\affiliation{$^1$School of Mathematics, University of Southampton, Southampton,
United Kingdom, SO17 1BJ}
\pacs{04.20.-q, 04.25.-g, 04.25.Nx, 04.30.Db}
\date{\today}

\begin{abstract}
Using a rigorous method of matched asymptotic expansions, I derive the equation
of motion of a small, compact body in an external vacuum spacetime through
second order in the body's mass (neglecting effects of internal structure). The
motion is found to be geodesic in a certain locally defined regular geometry
satisfying Einstein's equation at second order. I outline a method of
numerically obtaining both the metric of that regular geometry and the complete
second-order metric perturbation produced by the body. 
\end{abstract}
\maketitle

\emph{Introduction.} The governing equation of general relativity, the Einstein
field equation (EFE), describes how bodies influence spacetime curvature and
move within the resultant curved geometry. Yet, since the seminal work of
Einstein, Infeld, and Hoffman in 1938~\cite{Einstein-etal:38}, study of this
nonlinear problem of motion has largely focused on the post-Newtonian limit of
slow motion and weak fields. In the strong-field regime, bodies have instead
typically been approximated as test bodies moving in a spacetime that is
unaffected by them. Only within the last fifteen
years~\cite{Mino:96,Quinn-Wald:97} has there arisen an analytical description
of the gravitational backreaction: the body's perturbative effect on spacetime
geometry and that perturbation's effect on the body's motion. In the case of a
small mass $m$, this backreaction is called the gravitational self-force, and it
is now well understood at linear order in $m$~
\cite{Poisson:11,Gralla-Wald:08,Pound:10a,Pound:10b}. Beyond its foundational
role, the self-force is also potentially of great astrophysical importance, as
it describes the evolution of extreme-mass-ratio inspirals (EMRIs), in which a
stellar black hole or neutron star spirals into a supermassive black hole. Such
systems are predicted to be key sources for the planned gravitational wave
detector LISA~\cite{Gair:09}, and they will afford both a unique probe of
strong-field dynamics and a map of the spacetime near black holes. The
self-force also provides an essential point of comparison with other treatments
of the problem of motion: it complements post-Newtonian
theory~\cite{Favata:11,LeTiec-etal:11a} and fully nonlinear numerical
simulations~\cite{LeTiec-etal:11a,LeTiec-etal:11b}, both of which are ill-suited
to extreme mass ratios; and it fixes mass-dependent parameters in Effective One
Body (EOB) theory~\cite{Damour:09,Barack-Damour-Sago:10, Barausse-etal:11}.

However, to extract orbital parameters from a waveform emitted by an EMRI, one
requires a theoretical description accurate to second order in the small mass,
as shown by either naive or rigorous~\cite{Hinderer:08} scaling arguments.
Furthermore, comparisons with numerical simulations suggest that the
second-order self-force would provide a highly accurate description of
intermediate-mass-ratio binaries and even a reasonably accurate description of
similar-mass binaries~\cite{LeTiec-etal:11a,LeTiec-etal:11b}, both of which
should soon be observed by the ground-based detector Advanced
LIGO~\cite{Brown:07,Abadie:10}. The second-order force would also fix EOB
parameters quadratic in $m$. Although some work on the second-order problem has
been done~\cite{Rosenthal:06a,Rosenthal:06b}, it was performed in an
impractical gauge, with no clear means of calculating the force or the
perturbation producing it, and the basis of its approach was
problematic~\cite{Gralla-Wald:08,Pound:10a,Pound:10b,Pound:12}. Detweiler has
recently~\cite{Detweiler:11} examined the general features of the second-order
problem, but he made use of ill-defined equations and assumed (rather than derived) the equation of motion and a regularized second-order stress-energy tensor. Harte~\cite{Harte:11} 
has derived an equation of motion valid at all orders, but his derivation does not 
apply to the motion of black holes. None of these studies has provided a definite 
expression for, or means of calculating, the physical second-order metric perturbation or the piece of it that
determines the motion.

In this letter, I present the first complete description, explicitly determining
both the equation of motion and the metric perturbation. I use the
self-consistent formalism presented in~\cite{Pound:10a}. This formalism
incorporates the small body's finite size, involving no infinities or
regularization; it defines a worldline $\gamma$ that reflects the body's motion
(even a black hole's) on any timescale; it determines the equation of motion
directly from the EFE, with no further axioms; and it presents the perturbative
EFEs in a hyperbolic form convenient for numerical implementation. For
simplicity, I take the body to be spherical and nonspinning, neglecting higher
multipole moments. I work in units of $G=c=1$. Greek indices range from 0 to 3.
Lowercase Latin indices refer to spatial coordinates. Further details will
appear in a follow-up article~\cite{Pound:12}.

\emph{Self-consistent formalism.}
I combine two approximate solutions to the EFE, utilizing the method of matched
asymptotic expansions~\cite{Kevorkian-Cole:96, Eckhaus:79, Kates:81,
Pound:10b}. Suppose ${\sf g}_{\mu\nu}(\e)$ is an exact solution containing the
small body on a manifold $\mathcal{M}$, where $\e$ is an expansion parameter
that counts powers of the body's mass. Now let $r$ be some measure of distance
from the body and $\mathcal{R}$ represent the spacetime's lengthscales,
excluding those of the body itself. For $r\sim\mathcal{R}$, well outside the body, in any global
coordinates $x^\alpha$ in a vacuum region $\Omega$ (e.g., Boyer-Lindquist
coordinates of the supermassive black hole in an EMRI), I use the \emph{outer
expansion} ${\sf g}_{\mu\nu}(x^\alpha,\e)=g_{\mu\nu}(x^\alpha)+h_{\mu\nu}(x^\alpha,\e)$ on a
manifold $\mathcal{M}_E$. $(g_{\mu\nu},\mathcal{M}_E)$ defines an external
background spacetime with no small body in it, and
$h_{\mu\nu}(x,\e)=\sum_{n\geq1}\e^n h_{\alpha\beta}^{(n)}(x;\gamma(\e))$
describes perturbations due to the body, whose motion in $\mathcal{M}_E$ is
represented by $\gamma$. For $r\sim\e\mathcal{R}$, very near the body, the
metric varies rapidly, and there, in any coordinates $(T,X^a)$ approximately
centered on the body, I use the \emph{inner expansion} ${\sf
g}_{\mu\nu}(T,X^a/\e,\e)=g_{I\mu\nu}(T,X^a/\e)+\sum_{n\geq1}\e^n
H^{(n)}_{\mu\nu}(T,X^a/\e)$ on a manifold $\mathcal{M}_I$.
$(g_{I\mu\nu},\mathcal{M}_I)$ is the body's spacetime were it isolated, and
$H^{(n)}_{\mu\nu}$ describes perturbations due to interactions with the external
spacetime. The scaled coordinates $X^a/\e$ serve to keep the body's mass and
size fixed in the limit $\e\to0$, sending all other distances toward infinity;
the use of a single rescaling factor makes the approximation most appropriate
for compact bodies, in which the linear dimension is comparable to the mass (in
geometrized units). Scaling only distances, not $T$, is equivalent to assuming
the body possesses no fast internal dynamics.

In a \emph{buffer region} around the body, defined by $\e\ll r/\mathcal{R}\ll1$,
either expansion may be used, and since they approximate the same metric, they
must agree: the inner expansion can be expressed in unscaled coordinates and
then expanded for $r/\mathcal{R}\gg\e$, the outer expansion can be expanded for
$r/\mathcal{R}\ll1$, and the two results must match order by order in $r$ and
$\e$. It follows~\cite{Pound:10a} that $h^{(n)}_{\mu\nu}\sim1/r^n+O(r^{-n+1})$,
and the $1/r^n$ term is determined by the $(n-1)$th multipole moment of
$g_{I\mu\nu}$. If $r$ is a radial coordinate centered on $\gamma$ and this form
of $h^{(n)}_{\mu\nu}$ holds true, then both the body---in the full
spacetime---and $\gamma$---in the background spacetime---lie in the region
surrounded by the buffer. If all mass dipole terms also vanish in this
coordinate system, then the body is appropriately centered on $\gamma$.

In standard perturbation theory, the linearized EFE would constrain $\gamma$ to
be a geodesic in $g_{\mu\nu}$. To avoid this, I write the full, nonlinear EFE in $\Omega$ in
relaxed form by imposing the Lorenz gauge condition on the whole of
$h_{\mu\nu}$, rather than on each term $h^{(n)}_{\mu\nu}$:
\begin{equation}
g^{\nu\rho}\nabla_{\!\rho} \bar{h}_{\mu\nu}=0,\quad \text{where}\ 
\bar{h}_{\mu\nu}\equiv
h_{\mu\nu}-\frac{1}{2}g_{\mu\nu}g^{\rho\sigma}h_{\rho\sigma}.\label{gauge}
\end{equation}
The exact vacuum EFE in $\Omega$, ${\sf R}_{\mu\nu}=0$, then splits into a
sequence of wave equations,
\begin{align}
E_{\mu\nu}[h^{(1)}] &=0, \label{h1 eqn}\\
E_{\mu\nu}[h^{(2)}] &=2\delta^2 R_{\mu\nu}[h^{(1)}], \label{h2 eqn} 
\end{align}
etc., where $E_{\mu\nu}$ is the linear wave-operator $E_{\mu\nu}[h]=\Box
h_{\mu\nu}+2R_\mu{}^\rho{}_\nu{}^\sigma h_{\sigma\rho}$, $R_{\mu\nu\rho\sigma}$
is the Riemann tensor of the external background $g_{\mu\nu}$, and $\delta^2
R_{\mu\nu}$ is the part of the Ricci tensor quadratic in the metric
perturbation. No stress-energy tensor for the body appears here, since the body lies
outside $\Omega$. Equations~\eqref{h1 eqn} and \eqref{h2 eqn}
can be solved for arbitrary $\gamma$, the equation of motion of which is then
determined by the gauge condition. That condition will involve $\gamma$'s
$\e$-dependent acceleration $a^\mu(\tau,\e)$, where $\tau$ is proper time on
$\gamma$, and I split Eq.~\eqref{gauge} into a sequence of equations for each
$h^{(n)}_{\mu\nu}$ by substituting into it an expansion
$a^\mu=\sum_{n\geq0}{\e^n}a^{(n)\mu}$.

\emph{Outer expansion.}
In the buffer region, I construct the most general possible solution to the wave
equations \eqref{h1 eqn} and \eqref{h2 eqn} and gauge condition \eqref{gauge}.
That local solution is then used to construct a global solution. I work in
Fermi-Walker coordinates $(t,x^a)$ centered on $\gamma$, where $t$ coincides
with $\tau$ on $\gamma$ and $x^i$ are Cartesian coordinates on the spatial
submanifold transverse to $\gamma$ at time $t$. I assume each $h^{(n)}_{\mu\nu}$
can be expanded for small geodesic distance $r\equiv\sqrt{\delta_{ij}x^ix^j}$.
So at first order,
\begin{equation}
h^{(1)}_{\mu\nu}(t,x^a)=\sum_{m\geq-1,\ell\geq0}r^m h^{(1,m,\ell)}_{\mu\nu
L}(t)\nhat^L,\label{h1 form}
\end{equation}
where $L\equiv i_1\cdots i_\ell$, $n^i\equiv x^i/r$ is a radial unit vector, and
$\nhat^L\equiv n^{\langle i_1}\cdots n^{i_\ell\rangle}$ is the symmetric and
trace-free (STF) combination of $\ell$ such unit vectors; this decomposition in
terms of $\nhat^L$ is equivalent to an expansion in spherical
harmonics~\cite{Damour-Blanchet:86}. Substituting \eqref{h1 form} into \eqref{h1
eqn}, one finds that each term in Eq.~\eqref{h1 form} must satisfy a Poisson
equation, $\sum_{\ell}\partial^i\partial_i(r^m\nhat^L)h^{(1,m,\ell)}_{\mu\nu
L}(t)=r^{m-2}\sum_{\ell}S^{(1,m,\ell)}_{\mu\nu L}(t)\nhat^L$. The source
$S^{(1,m,\ell)}_{\mu\nu L}$ is a linear combination of the lower-order terms
$h^{(1,m'<m,\ell')}_{\mu\nu L'}$; it involves derivatives, $a^\mu$, and $R_{\mu\nu\rho\sigma}$. The general solution to this Poisson
equation consists of a homogeneous solution, comprising the single mode $h^{(1,m,\ell_H)}_{\mu\nu L_H}$ where $\ell_H=m$ for $m\geq0$ and $\ell_H=-m-1$ for $m<0$, plus a
particular inhomogeneous solution with modes $h^{(1,m,\ell){\rm P}}_{\mu\nu L}$ directly proportional to $S^{(1,m,\ell)}_{\mu\nu L}$. Given how $S^{(1,m,\ell)}_{\mu\nu L}$ is constructed, clearly each $h^{(1,m,\ell){\rm P}}_{\mu\nu L}$ is a linear combination of the lower-order modes $h^{(1,m'\leq m,\ell'_H)}_{\mu\nu L'_H}$; therefore, the general solution [of the form~\eqref{h1 form}] to Eq.~\eqref{h1 eqn} is wholly determined by the functions $h^{(1,m,\ell_H)}_{\mu\nu L_H}(t)$. The first of these, $h^{(1,-1,0)}_{\mu\nu}$, is fixed to be $2m\delta_{\mu\nu}$, where $m$ is the ADM mass of the inner background $g_{I\mu\nu}$~\cite{Gralla-Wald:08, Pound:10a}. All the others, $h^{(1,m\geq0,\ell_H)}_{\mu\nu L_H}$, are undetermined at this stage. It will thus prove convenient to split the general solution into two pieces: $h^{(1)}_{\mu\nu}=h^{(1){\rm S}}_{\mu\nu}+h^{(1){\rm R}}_{\mu\nu}$. I define the regular field
$h^{(1){\rm R}}_{\mu\nu}$ to comprise all terms involving the undetermined functions $h^{(1,m\geq0,\ell_H)}_{\mu\nu L_H}$. It reads
\begin{equation}
h^{(1){\rm R}}_{\mu\nu}=h^{(1,0,0)}_{\mu\nu}(t)+rh^{(1,1,1)}_{\mu\nu i}(t)n^i+O(r^2).\label{h1R}
\end{equation}
The singular field $h^{(1){\rm S}}_{\mu\nu}$ then comprises all the other terms; it is the particular solution obtained by setting $h^{(1,m\geq0,\ell_H)}_{\mu\nu L_H}=0$. It reads
\begin{align}
h^{(1){\rm S}}_{\mu\nu} &= \frac{2m}{r}\delta_{\mu\nu}+h^{(1,0,1)}_{\mu\nu i}(t)n^i\nonumber\\
&\quad+r\sum_{\ell=0,2}h^{(1,1,\ell)}_{\mu\nu L}(t)\nhat^L+O(r^2),\label{h1S}
\end{align}
where the functions $h^{(n,m,\ell)}_{\mu\nu L}(t)$ are linear in $m$. 
I stress that this split is merely a convenient grouping of terms in the general solution in the buffer region, with no impact on the results.
When so defined, $h^{(1){\rm S}}_{\mu\nu}$ and $h^{(1){\rm R}}_{\mu\nu}$ are each
solutions to the first-order field equations in $\Omega$. $h^{(1){\rm S}}_{\mu\nu}$ can be interpreted
as the body's bound field; it is determined solely by the fact that a compact body lies in the region surrounded by the buffer.
$h^{(1){\rm R}}_{\mu\nu}$ is a homogeneous (local) solution to the wave equation even at $r=0$,
propagating independently of the body; as a homogeneous
solution, it can be determined only by global boundary conditions. Imposing the
gauge condition determines the isotropic, $\delta_{\mu\nu}$ form of the $m/r$ term given above, as well as determining that the body behaves approximately as a
test particle, with constant mass (i.e., $\partial_tm=0$) and approximately
geodesic motion (i.e., $a^{(0)\mu}=0$).

One can prove with distributional~\cite{Gralla-Wald:08} or Green's
function~\cite{Pound:10a} methods that $h^{(1)}_{\mu\nu}$, because of its $m/r$
term, is sourced by the stress-energy tensor of a point particle,
$T^{(1)\mu\nu}=\int_{\gamma}mu^{\mu}u^\nu\frac{\delta^4(x^\alpha-z^\alpha(\tau))
}{\sqrt{-g}}d\tau$, where $z^\mu(\tau)$ is the parametrization of $\gamma$,
$u^\mu\equiv\frac{dz^\mu}{d\tau}$, and $g$ is the determinant of $g_{\mu\nu}$.
Therefore, at distances in the buffer region or greater, the body appears as a
point mass. With this determined, $h^{(1)}_{\mu\nu}$ can be found globally by
solving the wave equation
$E_{\mu\nu}[h^{(1)}]=-16\pi\overline{T}^{(1)}_{\mu\nu}$, where an overbar
indicates trace-reversal. Using retarded boundary conditions, for example, the
global solution will then fix the locally undetermined field $h^{(1){\rm
R}}_{\mu\nu}$. Numerous methods have been used to accomplish
this~\cite{Poisson:11}. In the case that $h^{(1)}_{\mu\nu}$ contains no
contribution from incoming waves at infinity, doing so
determines~\cite{Pound:10a} that at least through order $r$, $h^{(1){\rm
R}}_{\mu\nu}$ is the Detweiler-Whiting regular
field~\cite{Detweiler-Whiting:02}. 

The second-order solution proceeds almost identically. I assume an expansion
\begin{align}
h^{(2)}_{\mu\nu}(t,x^a) &= \sum_{m\geq-2,\ell\geq0}r^m h^{(2,m,\ell)}_{\mu\nu
L}(t)\nhat^L\nonumber\\
&\quad +\ln r\sum_{m\geq0,\ell\geq0}r^m h^{(2,m,\ln,\ell)}_{\mu\nu
L}(t)\nhat^L,\label{h2 form}
\end{align}
where the logarithmic terms arise from the correction $\sim m\ln(r/2m-1)$ to the
light cones in the buffer region (where $r\gg m$)~\cite{Pound:10a}. Substituting
this into Eq.~\eqref{h2 eqn}, together with an expansion of $h^{(1)}_{\mu\nu}$
up to order $r^2$, and finding the general solution at each order again
allows the convenient split $h^{(2)}_{\mu\nu}=h^{(2){\rm R}}_{\mu\nu}+h^{(2){\rm
S}}_{\mu\nu}$. However, here I define $h^{(2){\rm R}}_{\mu\nu}$ to comprise not
just all terms involving the undetermined functions $h^{(2,m\geq0,\ell_H)}_{\mu\nu L_H}(t)$, but also all terms quadratic in
$h^{(1){\rm R}}_{\mu\nu}$. This guarantees that $h^{(2){\rm R}}_{\mu\nu}$
satisfies $E_{\mu\nu}[h^{(2){\rm R}}]=2{\delta^2}R_{\mu\nu}[h^{(1){\rm R}}]$,
such that $g_{\mu\nu}+\e h^{(1){\rm R}}_{\mu\nu}+\e^2 h^{(2){\rm R}}_{\mu\nu}$
satisfies the vacuum EFE through order $\e^2$. Explicitly,
\begin{equation}
h^{(2){\rm R}}_{\mu\nu}=h^{(2,0,0)}_{\mu\nu}(t)+rh^{(2,1,1)}_{\mu\nu i}(t)n^i+O(r^2);\label{h2R}
\end{equation}
terms quadratic in $h^{(1){\rm R}}_{\mu\nu}$ would appear at order $r^2$. The singular field $h^{(2){\rm S}}_{\mu\nu}$, comprising all other terms in the general solution, then reads
\begin{align}
h^{(2){\rm S}}_{\mu\nu} &=
\frac{1}{r^2}\!\!\sum_{\ell=0,2}\!\!h^{(2,-2,\ell)}_{\mu\nu L}\nhat^L
+\frac{1}{r}\sum_{\ell=1}^3 h^{(2,-1,\ell)}_{\mu\nu L}\nhat^L+\frac{2\delta
m_{\mu\nu}}{r}\nonumber\\
&\quad +\sum_{\ell=1}^4h^{(2,0,\ell)}_{\mu\nu
L}\nhat^L+r\!\!\!\!\!\!\!\sum_{\ell=0,2,3,4,5}\!\!\!\!\! h^{(2,0,\ell)}_{\mu\nu
L}\nhat^L\nonumber\\
&\quad +\ln r \big[h^{(2,0,\ln,0)}_{\mu\nu}+rh^{(2,1,\ln,1)}_{\mu\nu
i}n^i\big]+O(r^2),\label{h2S}
\end{align}
where $\delta m_{\mu\nu}$ is a mass-like tensor defined on $\gamma$. The explicit terms in this expansion can be found in Ref.~\cite{Pound:10a} through order $r^0$ when $a^\mu=0$, and they will be written out in full in the follow-up
article~\cite{Pound:12}; the functions $h^{(2,m,\ell)}_{\mu\nu L}(t)$ are linear
combinations of $m^2$, $mh^{(1){\rm R}}_{\mu\nu}$, and $\delta m_{\mu\nu}$. The body's dipoles (i.e., those of $g_{I\mu\nu}$) would contribute to $h^{(2){\rm S}}_{\mu\nu}$, but I set the spin
to zero for simplicity and the mass dipole to zero to ensure that $\gamma$
accurately represents the body's motion. Note that my definition of $h^{(2){\rm R}}_{\mu\nu}$ means that $h^{(2){\rm
S}}_{\mu\nu}$ satisfies not Eq.~\eqref{h2 eqn}, but $E_{\mu\nu}[h^{(2){\rm
S}}]=2{\delta^2}R_{\mu\nu}[h^{(1)}]-2{\delta^2}R_{\mu\nu}[h^{(1){\rm R}}]$.

As at first order, the gauge condition determines the form of $\delta
m_{\mu\nu}$, as given in Table~\ref{delta_m}, along with the first-order
acceleration, 
\begin{equation}
a^{(1)}_i(t)=\frac{1}{2}h^{(1,1,1)}_{tti}(t)\big|_{a^\mu=0}-h^{(1,0,0)}_{ti,t}
(t)\big|_{a^\mu=0}.\label{a1}
\end{equation}
The evaluation at $a^\mu=0$ is to be performed only at time $t$, leaving the
past history of $\gamma$ unchanged; this follows from the presumed expansion of
$a^\mu$, and it prevents a need for order-reduction~\cite{Pound:10a,Poisson:11}.
One can show Eq.~\eqref{a1} is equivalent to the geodesic equation in
$g_{\mu\nu}+\e h^{(1){\rm R}}_{\mu\nu}$ at order $\e$~\cite{Poisson:11}.

\begin{table}
\caption{\label{delta_m}Components of $\delta m_{\mu\nu}(t)$ in terms of the
first-order regular field $h^{(1){\rm
R}}_{\mu\nu}(t,r=0)= h^{(1,0,0)}_{\mu\nu}(t)$.}
\begin{ruledtabular}
\begin{tabular}{l}
$\delta m_{tt}=-m h^{(1,0,0)}_{tt}-\frac{1}{6}m\delta^{ij}h^{(1,0,0)}_{ij}$\\
$\delta m_{ta}=-\frac{2}{3}mh^{(1,0,0)}_{ta}$\\
$\delta m_{ab}=\delta_{ab}\left(\frac{1}{3}mh^{(1,0,0)}_{tt}+\frac{5}{18}m\delta^{ij}
h^{(1,0,0)}_{ij}\right)+\frac{1}{3}mh^{(1,0,0)}_{\langle ab\rangle}$
\end{tabular}
\end{ruledtabular}
\end{table}

Using the same Green's-function method \cite{Pound:10a} as at first order, one
can straightforwardly prove that the terms involving $\delta m_{\mu\nu}$ have a
point source, the trace-reversal of which is given by the effective
stress-energy tensor 
\begin{equation}
T^{(2)}_{\mu\nu}=\int_\gamma \frac{1}{2}\overline{\delta
m}_{\mu\nu}(\tau)\frac{\delta^4(x^\alpha-z^\alpha(\tau))}{\sqrt{-g}}d\tau.
\end{equation}
The global solution to the second-order EFE sourced by the body can then be
obtained numerically via a puncture scheme~\cite{Dolan:11}: Outside a tube $\Gamma$ around
the body, one can solve Eq.~\eqref{h2 eqn} directly; inside the tube, one can
use an approximation $\tilde h^{(2){\rm S}}_{\mu\nu}$ to $h^{(2){\rm
S}}_{\mu\nu}$, given by Eq.~\eqref{h2S} without the ``$O(r^2)$" term, and a
regular field $\tilde h^{(2){\rm R}}_{\mu\nu}\equiv h^{(2)}_{\mu\nu}-\tilde
h^{(2){\rm S}}_{\mu\nu}$, which satisfies
\begin{equation}
E_{\mu\nu}[\tilde h^{(2){\rm R}}]=-16\pi\overline{T}^{(2)}_{\mu\nu}+2\delta^2R_{\mu\nu}[h^{(1)}]-E_{\mu\nu}[
\tilde h^{(2){\rm S}}].\label{puncture}
\end{equation}
All divergent terms on the right-hand side cancel, leaving $\tilde h^{(2){\rm
R}}_{\mu\nu}$ to solve a wave equation with a regular source. At $\Gamma$, the
analytical expression for $\tilde h^{(2){\rm S}}_{\mu\nu}$ can be added to the numerical solution to Eq.~\eqref{puncture} to change variables to the full field. Since $\tilde h^{(2){\rm R}}_{\mu\nu}$ will agree with $h^{(2){\rm
R}}_{\mu\nu}$ through order $r$, this procedure will also determine $h^{(2,0,0)}_{\mu\nu}(t)$ and $h^{(2,1,1)}_{\mu\nu i}(t)$, which, as we shall find. This puncture scheme
can be implemented immediately after transforming Eq.~\eqref{h2S} to a
desired coordinate system.

\emph{Matching to an inner expansion.}
One could proceed with the same method to find $h^{(3)}_{\mu\nu}$ together with
$a^{(2)\mu}$. I instead take a more efficient route by determining $a^{(2)\mu}$
from additional information about the inner expansion. I take the small body to
be a Schwarzschild black hole. Since the inner expansion affects the outer one
solely through the body's multipole moments, this amounts to neglecting
non-monopole moments, as was done in the preceding section. Beyond effects of
those moments, the equation of motion I derive will hold for any compact body. I
also specify the perturbations to be produced by tidal fields. These fields are
of quadrupole order and higher, and while they produce mass and spin
perturbations via tidal heating and torquing, dimensional analysis shows those
perturbations do not contribute at the orders in $\e$ of interest. With these
specifications, I follow the procedure in Ref.~\cite{Pound:10b}, insisting that
in a suitable mass-centered coordinate system, this inner expansion must equal
the outer expansion in Fermi-Walker coordinates when expanded in the buffer
region, up to a unique gauge transformation that excludes spatial translations
at $\gamma$; this ensures the desired relationship between $\gamma$ and the
mass-centered inner expansion. Here `mass-centered coordinate system' means one
in which the mass dipole of $g_{I\mu\nu}$ vanishes along with all even-parity
dipole perturbations that behave as a mass dipole, scaling as $1/r^2$ in the buffer
region. Other even-parity dipole perturbations are also set to zero in order to
leave no residual gauge freedom. Because the Fermi-Walker coordinates are
$\e$-dependent, each term in the outer expansion depends on $a^\mu$. Hence,
prior to matching the metrics, I substitute $a^\mu=\sum_{n\geq0}\e^na_{(n)}^\mu$
into $g_{\mu\nu}+\sum_{n\geq1}\e^n h^{(n)}_{\mu\nu}$.

A tidally perturbed Schwarzschild metric is given in Ref.~\cite{Poisson:05} in
advanced Eddington-Finkelstein coordinates and a light-cone gauge. The tidal
fields are represented by STF tensorial functions of time: $\E_{ij}$ and
$\B_{ij}$ for electric- and magnetic-type quadrupole tides and $\E_{ijk}$ and
$\B_{ijk}$ for analogous octupole tides; for the powers of $r$ of interest,
hexadecapole and higher tidal fiels would appear only at order $\e^3$ in the
outer expansion. Transforming to a suitable Fermi-like coordinate system and
expanding the result to order $\e^2$ in the buffer region yields
\begin{align}
{\sf g}_{tt} &=
-f_{tt}+r^2H^{(2,2)}_{tti_1i_2}\nhat^{i_1i_2}+r^3\sum_{\ell=2,3}H^{(3,\ell)}_{
ttL}\nhat^L \label{inner tt}\\
{\sf g}_{ta} &=
r^2\sum_{\ell=1}^3H^{(2,\ell)}_{taL}\nhat^L+r^3\sum_{\ell=1}^4H^{(3,\ell)}_{taL}
\nhat^L\\
{\sf g}_{ab} &=
f_{ab}\!+\!f\nhat_{ab}\!+\!r^2\!\sum_{\ell=0}^4H^{(2,\ell)}_{abL}
\nhat^L\!+\!r^3\!\sum_{\ell=0}^5H^{(3,\ell)}_{abL}\nhat^L.\label{inner ab}
\end{align}
Here $f_{tt}=1-\frac{2\e{m}}{r}+\frac{2\e^2{m}^2}{r^2}$, $f_{ab}=(1+\frac{2\e{m}}{r}
+\frac{4\e^2{m}^2}{3r^2})\delta_{ab}$, and $f=\frac{\e^2{m}^2}{r^2}$ describe
the Schwarzschild metric in harmonic coordinates. The coefficients
$H^{(2,\ell)}_{\mu\nu L}$ are functions of $\e{m}/r$ forming linear combinations
of $\E_{ij}(t)$ and $\B_{ij}(t)$, and $H^{(3,\ell)}_{\mu\nu L}$ are linear
combinations of $\dot\E_{ij}(t)$, $\dot\B_{ij}(t)$, $\E_{ijk}(t)$, and
$\B_{ijk}(t)$, where an overdot indicates a time-derivative. For example,
$H^{(2,2)}_{ttij}=-(1-\frac{5\e{m}}{3r}+\frac{4\e^2{m}^2}{3r^2})\E_{ij}$.

If explicit appearances of $\mu$ are set to zero, the metric of Eqs.~\eqref{inner
tt}--\eqref{inner ab} reduces to that of a vaccum spacetime in Fermi coordinates
centered on a geodesic, with $\E_{ab}=\E^{(0)}_{ab}+\e\delta\E_{ab}+O(\e^2)$,
$\B_{ab}=\B^{(0)}_{ab}+\e\delta\B_{ab}+O(\e^2)$,
$\E_{abc}=\E^{(0)}_{abc}+O(\e)$, and $\B_{abc}=\B^{(0)}_{abc}+O(\e)$, where the
zeroth-order fields $\E^{(0)}_{ab}$, $\B^{(0)}_{ab}$, etc., are components of
$R_{\mu\nu\rho\sigma}$ and its first derivative evaluated at $r=0$. There is no
manifest appearance of the fields $h^{(n){\rm R}}_{\mu\nu}$ in Eqs.~\eqref{inner
tt}--\eqref{inner ab}; they are incorporated into the corrections
$\delta\E_{ab}$, $\delta\B_{ab}$, etc. More significantly, there is no term
corresponding to an acceleration; any such term would induce a mass-dipole-like
term and would vanish in mass-centered coordinates.

To match the expansions at orders $\e$ and $\e^2$, I seek a unique
transformation $x^\mu\to x^\mu-\e \xi^{(1)\mu}-\e^2\xi^{(2)\mu}$ that brings the
outer expansion into the form of Eqs.~\eqref{inner tt}--\eqref{inner ab}.
Decomposing $\xi^{(1)\mu}$ and $\xi^{(2)\mu}$ into irreducible STF pieces, one
readily finds a unique transformation. At each order in $\e$, this
transformation can be thought of as putting the outer expansion into Fermi
coordinates in $g_{\mu\nu}+\e h^{\rm R}_{\mu\nu}$. The order-$\e$ transformation
is given in Ref.~\cite{Pound:10b} up to order-$r$ terms in the metric; matching
the metrics exhausts all freedom in that transformation and uniquely determines
the standard result~\eqref{a1} for $a^{(1)\mu}$. Matching order-$\e r^2$ terms
in the metric fixes $\delta\E_{ij}$ and $\delta\B_{ij}$ in terms of
$h^{(1)}_{\mu\nu}$. The order-$\e^2$ transformation, when carried to order-$r$
terms in the metric, likewise uniquely determines 
\begin{align}
a_i^{(2)} &=
\frac{1}{2}h^{(2,1,1)}_{tti}\big|_{a^\mu=0}-h^{(2,0,0)}_{ti,t}\big|_{a^\mu=0}
+h^{(1,0,0)}_{tt}a^{(1)}_i\nonumber\\
&\quad-\frac{1}{2}h^{(1,0,0)}_{ti}h^{(1,0,0)}_{tt,t}-\frac{11}{3}m\dot
a^{(1)}_i,\label{a2}
\end{align}
where, as in Eq.~\eqref{a1}, the evaluation at $a^\mu=0$ occurs only at time
$t$. Summing $\e a^\mu_{(1)}$ and $\e^2a^\mu_{(2)}$, one finds, up to $O(\e^3)$
errors,
\begin{align}
a^\mu\!=\!\frac{1}{2}\!\left(g^{\mu\nu}\!\!+\!u^\mu
u^\nu\right)\!\!\left(g_\nu{}^\rho\!-\!h^{{\rm
R}}_\nu{}^\rho\right)\!\!\left(h^{\rm R}_{\sigma\lambda;\rho}\!-\!2h^{\rm
R}_{\rho\sigma;\lambda}\right)\!\!u^\sigma\! u^\lambda,\label{acceleration}
\end{align}
where $h^{\rm R}_{\mu\nu}=\e h^{(1){\rm R}}_{\mu\nu}+\e^2h^{(2){\rm
R}}_{\mu\nu}$. This is the geodesic equation in the locally defined regular
metric $g_{\mu\nu}+h^{\rm R}_{\mu\nu}$, up to terms cubic in $h^{\rm
R}_{\mu\nu}$. $g_{\mu\nu}+h^{\rm R}_{\mu\nu}$ may trivially be extended to a
$C^n$ (local) vacuum solution to the EFE, through order $\e^2$, by finding the
solution \eqref{h2 form} through order $r^n$.

Equation~\eqref{acceleration} agrees with the form of Harte's equation of
motion~\cite{Harte:11} but represents a major advance: it has been shown to
apply to black holes, and it comes along with a concrete means of calculating
both $a^{(2)\mu}$ and $h^{(2)}_{\mu\nu}$. 

\emph{Discussion.}
I have shown that through second order in its mass, a small body moves on a
geodesic of a certain locally defined regular metric. I have also derived
results, given by \eqref{h2S}, \eqref{puncture}, and \eqref{a2}, that (together with the first-order equations) may be used to simultaneously evolve the body's position and find the perturbation due to it, thereby solving the EFE through second order. Although these results were derived only for a nonrotating
black hole, they should hold for any spherical, compact body with slow internal
dynamics. For nonspherical bodies, they will be modified by higher multipole
moments: $h^{(2){\rm R}}_{\mu\nu}$ and $h^{(2){\rm S}}_{\mu\nu}$ will be straightforwardly altered
by the body's spin~\cite{Pound:12}, and $a^{(2)\mu}$
will include well-known~\cite{Dixon:74,Steinhoff:10,Harte:11} couplings of the
moments to the external curvature.

\begin{acknowledgments}
I wish to thank Leor Barack for many helpful discussions and suggested
improvements to this manuscript. This work was supported by the Natural Sciences
and Engineering Research Council of Canada.
\end{acknowledgments}

\bibliography{prl}

\begin{thebibliography}{10}%
\makeatletter
\providecommand \@ifxundefined [1]{%
 \ifx #1\undefined \expandafter \@firstoftwo
 \else \expandafter \@secondoftwo
\fi
}%
\providecommand \@ifnum [1]{%
 \ifnum #1\expandafter \@firstoftwo
 \else \expandafter \@secondoftwo
\fi
}%
\providecommand \enquote [1]{``#1''}%
\providecommand \bibnamefont  [1]{#1}%
\providecommand \bibfnamefont [1]{#1}%
\providecommand \citenamefont [1]{#1}%
\providecommand\href[0]{\@sanitize\@href}%
\providecommand\@href[1]{\endgroup\@@startlink{#1}\endgroup\@@href}%
\providecommand\@@href[1]{#1\@@endlink}%
\providecommand \@sanitize [0]{\begingroup\catcode`\&12\catcode`\#12\relax}%
\@ifxundefined \pdfoutput {\@firstoftwo}{%
 \@ifnum{\z@=\pdfoutput}{\@firstoftwo}{\@secondoftwo}%
}{%
 \providecommand\@@startlink[1]{\leavevmode\special{html:<a href="#1">}}%
 \providecommand\@@endlink[0]{\special{html:</a>}}%
}{%
 \providecommand\@@startlink[1]{%
  \leavevmode
  \pdfstartlink
   attr{/Border[0 0 1 ]/H/I/C[0 1 1]}%
   user{/Subtype/Link/A<</Type/Action/S/URI/URI(#1)>>}%
  \relax
 }%
 \providecommand\@@endlink[0]{\pdfendlink}%
}%
\providecommand \url  [0]{\begingroup\@sanitize \@url }%
\providecommand \@url [1]{\endgroup\@href {#1}{\urlprefix}}%
\providecommand \urlprefix [0]{URL }%
\providecommand \Eprint[0]{\href }%
\@ifxundefined \urlstyle {%
  \providecommand \doi [1]{doi:\discretionary{}{}{}#1}%
}{%
  \providecommand \doi [0]{doi:\discretionary{}{}{}\begingroup
  \urlstyle{rm}\Url }%
}%
\providecommand \doibase [0]{http://dx.doi.org/}%
\providecommand \Doi[1]{\href{\doibase#1}}%
\providecommand \bibAnnote [3]{%
  \BibitemShut{#1}%
  \begin{quotation}\noindent
    \textsc{Key:}\ #2\\\textsc{Annotation:}\ #3%
  \end{quotation}%
}%
\providecommand \bibAnnoteFile [2]{%
  \IfFileExists{#2}{\bibAnnote {#1} {#2} {\input{#2}}}{}%
}%
\providecommand \typeout [0]{\immediate \write \m@ne }%
\providecommand \selectlanguage [0]{\@gobble}%
\providecommand \bibinfo [0]{\@secondoftwo}%
\providecommand \bibfield [0]{\@secondoftwo}%
\providecommand \translation [1]{[#1]}%
\providecommand \BibitemOpen[0]{}%
\providecommand \bibitemStop [0]{}%
\providecommand \bibitemNoStop [0]{.\EOS\space}%
\providecommand \EOS [0]{\spacefactor3000\relax}%
\providecommand \BibitemShut [1]{\csname bibitem#1\endcsname}%
\bibitem{Einstein-etal:38}%
  \BibitemOpen
  \bibfield{author}{%
  \bibinfo {author} {\bibfnamefont{A.}~\bibnamefont{Einstein}}, \bibinfo
  {author} {\bibfnamefont{L.}~\bibnamefont{Infeld}},\ and\ \bibinfo {author}
  {\bibfnamefont{B.}~\bibnamefont{Hoffmann}},\ }%
  \bibfield{journal}{%
  \bibinfo {journal} {Annals of Mathematics}\ }%
  \textbf{\bibinfo {volume} {39}},\ \bibinfo {pages} {65} (\bibinfo {year}
  {1938})%
  \bibAnnoteFile{NoStop}{Einstein-etal:38}%
\bibitem{Mino:96}%
  \BibitemOpen
  \bibfield{author}{%
  \bibinfo {author} {\bibfnamefont{Y.}~\bibnamefont{Mino}}, \bibinfo {author}
  {\bibfnamefont{M.}~\bibnamefont{Sasaki}},\ and\ \bibinfo {author}
  {\bibfnamefont{T.}~\bibnamefont{Tanaka}},\ }%
  \bibfield{journal}{%
  \bibinfo {journal} {Phys.Rev.}\ }%
  \textbf{\bibinfo {volume} {D55}},\ \bibinfo {pages} {3457} (\bibinfo {year}
  {1997})%
  \bibAnnoteFile{NoStop}{Mino:96}%
\bibitem{Quinn-Wald:97}%
  \BibitemOpen
  \bibfield{author}{%
  \bibinfo {author} {\bibfnamefont{T.~C.}\ \bibnamefont{Quinn}}\ and\ \bibinfo
  {author} {\bibfnamefont{R.~M.}\ \bibnamefont{Wald}},\ }%
  \bibfield{journal}{%
  \bibinfo {journal} {Phys.Rev.}\ }%
  \textbf{\bibinfo {volume} {D56}},\ \bibinfo {pages} {3381} (\bibinfo {year}
  {1997})%
  \bibAnnoteFile{NoStop}{Quinn-Wald:97}%
\bibitem{Poisson:11}%
  \BibitemOpen
  \bibfield{author}{%
  \bibinfo {author} {\bibfnamefont{E.}~\bibnamefont{Poisson}}, \bibinfo
  {author} {\bibfnamefont{A.}~\bibnamefont{Pound}},\ and\ \bibinfo {author}
  {\bibfnamefont{I.}~\bibnamefont{Vega}},\ }%
  \bibfield{journal}{%
  \bibinfo {journal} {Living Rev. Relativity}\ }%
  \textbf{\bibinfo {volume} {14}} (\bibinfo {year} {2011}),\
  \url{http://www.livingreviews.org/lrr-2011-7}%
  \bibAnnoteFile{NoStop}{Poisson:11}%
\bibitem{Gralla-Wald:08}%
  \BibitemOpen
  \bibfield{author}{%
  \bibinfo {author} {\bibfnamefont{S.~E.}\ \bibnamefont{Gralla}}\ and\ \bibinfo
  {author} {\bibfnamefont{R.~M.}\ \bibnamefont{Wald}},\ }%
  \bibfield{journal}{%
  \bibinfo {journal} {Class.Quant.Grav.}\ }%
  \textbf{\bibinfo {volume} {25}},\ \bibinfo {pages} {205009} (\bibinfo {year}
  {2008})%
  \bibAnnoteFile{NoStop}{Gralla-Wald:08}%
\bibitem{Pound:10a}%
  \BibitemOpen
  \bibfield{author}{%
  \bibinfo {author} {\bibfnamefont{A.}~\bibnamefont{Pound}},\ }%
  \bibfield{journal}{%
  \bibinfo {journal} {Phys. Rev. D}\ }%
  \textbf{\bibinfo {volume} {81}},\ \bibinfo {pages} {024023} (\bibinfo {year}
  {2010})%
  \bibAnnoteFile{NoStop}{Pound:10a}%
\bibitem{Pound:10b}%
  \BibitemOpen
  \bibfield{author}{%
  \bibinfo {author} {\bibfnamefont{A.}~\bibnamefont{Pound}},\ }%
  \bibfield{journal}{%
  \bibinfo {journal} {Phys. Rev. D}\ }%
  \textbf{\bibinfo {volume} {81}},\ \bibinfo {pages} {124009} (\bibinfo {year}
  {2010})%
  \bibAnnoteFile{NoStop}{Pound:10b}%
\bibitem{Gair:09}%
  \BibitemOpen
  \bibfield{author}{%
  \bibinfo {author} {\bibfnamefont{J.~R.}\ \bibnamefont{Gair}},\ }%
  \bibfield{journal}{%
  \bibinfo {journal} {Class.Quant.Grav.}\ }%
  \textbf{\bibinfo {volume} {26}},\ \bibinfo {pages} {094034} (\bibinfo {year}
  {2009})%
  \bibAnnoteFile{NoStop}{Gair:09}%
\bibitem{Favata:11}%
  \BibitemOpen
  \bibfield{author}{%
  \bibinfo {author} {\bibfnamefont{M.}~\bibnamefont{Favata}},\ }%
  \bibfield{journal}{%
  \bibinfo {journal} {Phys.Rev.}\ }%
  \textbf{\bibinfo {volume} {D83}},\ \bibinfo {pages} {024028} (\bibinfo {year}
  {2011})%
  \bibAnnoteFile{NoStop}{Favata:11}%
\bibitem{LeTiec-etal:11a}%
  \BibitemOpen
  \bibfield{author}{%
  \bibinfo {author} {\bibfnamefont{A.}~\bibnamefont{Le~Tiec}}, \bibinfo
  {author} {\bibfnamefont{A.~H.}\ \bibnamefont{Mroue}}, \bibinfo {author}
  {\bibfnamefont{L.}~\bibnamefont{Barack}}, \bibinfo {author}
  {\bibfnamefont{A.}~\bibnamefont{Buonanno}}, \bibinfo {author}
  {\bibfnamefont{H.~P.}\ \bibnamefont{Pfeiffer}}, \emph{et~al.},\ }%
  \bibfield{journal}{%
  \bibinfo {journal} {Phys.Rev.Lett.}\ }%
  \textbf{\bibinfo {volume} {107}},\ \bibinfo {pages} {141101} (\bibinfo {year}
  {2011})%
  \bibAnnoteFile{NoStop}{LeTiec-etal:11a}%
\bibitem{LeTiec-etal:11b}%
  \BibitemOpen
  \bibfield{author}{%
  \bibinfo {author} {\bibfnamefont{A.}~\bibnamefont{Le~Tiec}}, \bibinfo
  {author} {\bibfnamefont{E.}~\bibnamefont{Barausse}},\ and\ \bibinfo {author}
  {\bibfnamefont{A.}~\bibnamefont{Buonanno}}}%
   (\bibinfo {year} {2011}),\
  \Eprint{http://arxiv.org/abs/1111.5609}{arXiv:1111.5609}%
  \bibAnnoteFile{NoStop}{LeTiec-etal:11b}%
\bibitem{Damour:09}%
  \BibitemOpen
  \bibfield{author}{%
  \bibinfo {author} {\bibfnamefont{T.}~\bibnamefont{Damour}},\ }%
  \bibfield{journal}{%
  \bibinfo {journal} {Phys.Rev.}\ }%
  \textbf{\bibinfo {volume} {D81}},\ \bibinfo {pages} {024017} (\bibinfo {year}
  {2010})%
  \bibAnnoteFile{NoStop}{Damour:09}%
\bibitem{Barack-Damour-Sago:10}%
  \BibitemOpen
  \bibfield{author}{%
  \bibinfo {author} {\bibfnamefont{L.}~\bibnamefont{Barack}}, \bibinfo {author}
  {\bibfnamefont{T.}~\bibnamefont{Damour}},\ and\ \bibinfo {author}
  {\bibfnamefont{N.}~\bibnamefont{Sago}},\ }%
  \bibfield{journal}{%
  \bibinfo {journal} {Phys.Rev.}\ }%
  \textbf{\bibinfo {volume} {D82}},\ \bibinfo {pages} {084036} (\bibinfo {year}
  {2010})%
  \bibAnnoteFile{NoStop}{Barack-Damour-Sago:10}%
\bibitem{Barausse-etal:11}%
  \BibitemOpen
  \bibfield{author}{%
  \bibinfo {author} {\bibfnamefont{E.}~\bibnamefont{Barausse}}, \bibinfo
  {author} {\bibfnamefont{A.}~\bibnamefont{Buonanno}},\ and\ \bibinfo {author}
  {\bibfnamefont{A.}~\bibnamefont{Le~Tiec}}}%
   (\bibinfo {year} {2011}),\
  \Eprint{http://arxiv.org/abs/1111.5610}{arXiv:1111.5610 [gr-qc]}%
  \bibAnnoteFile{NoStop}{Barausse-etal:11}%
\bibitem{Hinderer:08}%
  \BibitemOpen
  \bibfield{author}{%
  \bibinfo {author} {\bibfnamefont{T.}~\bibnamefont{Hinderer}}\ and\ \bibinfo
  {author} {\bibfnamefont{E.~E.}\ \bibnamefont{Flanagan}},\ }%
  \bibfield{journal}{%
  \bibinfo {journal} {Phys.Rev.}\ }%
  \textbf{\bibinfo {volume} {D78}},\ \bibinfo {pages} {064028} (\bibinfo {year}
  {2008})%
  \bibAnnoteFile{NoStop}{Hinderer:08}%
\bibitem{Brown:07}%
  \BibitemOpen
  \bibfield{author}{%
  \bibinfo {author} {\bibfnamefont{D.~A.}\ \bibnamefont{Brown}}, \bibinfo
  {author} {\bibfnamefont{J.}~\bibnamefont{Brink}}, \bibinfo {author}
  {\bibfnamefont{H.}~\bibnamefont{Fang}}, \bibinfo {author}
  {\bibfnamefont{J.~R.}\ \bibnamefont{Gair}}, \bibinfo {author}
  {\bibfnamefont{C.}~\bibnamefont{Li}}, \bibinfo {author}
  {\bibfnamefont{G.}~\bibnamefont{Lovelace}}, \bibinfo {author}
  {\bibfnamefont{I.}~\bibnamefont{Mandel}},\ and\ \bibinfo {author}
  {\bibfnamefont{K.~S.}\ \bibnamefont{Thorne}},\ }%
  \bibfield{journal}{%
  \bibinfo {journal} {Phys. Rev. Lett.}\ }%
  \textbf{\bibinfo {volume} {99}},\ \bibinfo {pages} {201102} (\bibinfo {year}
  {2007})%
  \bibAnnoteFile{NoStop}{Brown:07}%
\bibitem{Abadie:10}%
  \BibitemOpen
  \bibfield{author}{%
  \bibinfo {author} {\bibfnamefont{J.}~\bibnamefont{Abadie}} \emph{et~al.}
  (\bibinfo {collaboration} {LIGO Scientific Collaboration, Virgo
  Collaboration}),\ }%
  \bibfield{journal}{%
  \bibinfo {journal} {Class.Quant.Grav.}\ }%
  \textbf{\bibinfo {volume} {27}},\ \bibinfo {pages} {173001} (\bibinfo {year}
  {2010})%
  \bibAnnoteFile{NoStop}{Abadie:10}%
\bibitem{Rosenthal:06a}%
  \BibitemOpen
  \bibfield{author}{%
  \bibinfo {author} {\bibfnamefont{E.}~\bibnamefont{Rosenthal}},\ }%
  \bibfield{journal}{%
  \bibinfo {journal} {Phys.Rev.}\ }%
  \textbf{\bibinfo {volume} {D73}},\ \bibinfo {pages} {044034} (\bibinfo {year}
  {2006})%
  \bibAnnoteFile{NoStop}{Rosenthal:06a}%
\bibitem{Rosenthal:06b}%
  \BibitemOpen
  \bibfield{author}{%
  \bibinfo {author} {\bibfnamefont{E.}~\bibnamefont{Rosenthal}},\ }%
  \bibfield{journal}{%
  \bibinfo {journal} {Phys.Rev.}\ }%
  \textbf{\bibinfo {volume} {D74}},\ \bibinfo {pages} {084018} (\bibinfo {year}
  {2006})%
  \bibAnnoteFile{NoStop}{Rosenthal:06b}%
\bibitem{Pound:12}%
  \BibitemOpen
  \bibfield{author}{%
  \bibinfo {author} {\bibfnamefont{A.}~\bibnamefont{Pound}}}%
   (\bibinfo {year} {in preparation})%
  \bibAnnoteFile{NoStop}{Pound:12}%
\bibitem{Detweiler:11}%
  \BibitemOpen
  \bibfield{author}{%
  \bibinfo {author} {\bibfnamefont{S.}~\bibnamefont{Detweiler}}}%
   (\bibinfo {year} {2011}),\
  \Eprint{http://arxiv.org/abs/1107.2098}{arXiv:1107.2098 [gr-qc]}%
  \bibAnnoteFile{NoStop}{Detweiler:11}%
\bibitem{Harte:11}%
  \BibitemOpen
  \bibfield{author}{%
  \bibinfo {author} {\bibfnamefont{A.~I.}\ \bibnamefont{Harte}}}%
   (\bibinfo {year} {2011}),\
  \Eprint{http://arxiv.org/abs/1103.0543}{arXiv:1103.0543 [gr-qc]}%
  \bibAnnoteFile{NoStop}{Harte:11}%
\bibitem{Kevorkian-Cole:96}%
  \BibitemOpen
  \bibfield{author}{%
  \bibinfo {author} {\bibfnamefont{J.}~\bibnamefont{Kevorkian}}\ and\ \bibinfo
  {author} {\bibfnamefont{J.~D.}\ \bibnamefont{Cole}},\ }%
  \emph{\bibinfo {title} {Multiple scale and singular perturbation methods}}\
  (\bibinfo {publisher} {Springer},\ \bibinfo {address} {New York},\ \bibinfo
  {year} {1996})%
  \bibAnnoteFile{NoStop}{Kevorkian-Cole:96}%
\bibitem{Eckhaus:79}%
  \BibitemOpen
  \bibfield{author}{%
  \bibinfo {author} {\bibfnamefont{W.}~\bibnamefont{Eckhaus}},\ }%
  \emph{\bibinfo {title} {Asymptotic Analysis of Singular Perturbations}}\
  (\bibinfo {publisher} {Elsevier North-Holland},\ \bibinfo {address} {New
  York},\ \bibinfo {year} {1979})%
  \bibAnnoteFile{NoStop}{Eckhaus:79}%
\bibitem{Kates:81}%
  \BibitemOpen
  \bibfield{author}{%
  \bibinfo {author} {\bibfnamefont{R.}~\bibnamefont{Kates}},\ }%
  \bibfield{journal}{%
  \bibinfo {journal} {Ann. Phys. (N.Y.)}\ }%
  \textbf{\bibinfo {volume} {132}},\ \bibinfo {pages} {1} (\bibinfo {year}
  {1981})%
  \bibAnnoteFile{NoStop}{Kates:81}%
\bibitem{Damour-Blanchet:86}%
  \BibitemOpen
  \bibfield{author}{%
  \bibinfo {author} {\bibfnamefont{T.}~\bibnamefont{Damour}}\ and\ \bibinfo
  {author} {\bibfnamefont{L.}~\bibnamefont{Blanchet}},\ }%
  \bibfield{journal}{%
  \bibinfo {journal} {Phil. Trans. R. Soc. Lond. A}\ }%
  \textbf{\bibinfo {volume} {320}},\ \bibinfo {pages} {379} (\bibinfo {year}
  {1986})%
  \bibAnnoteFile{NoStop}{Damour-Blanchet:86}%
\bibitem{Detweiler-Whiting:02}%
  \BibitemOpen
  \bibfield{author}{%
  \bibinfo {author} {\bibfnamefont{S.~L.}\ \bibnamefont{Detweiler}}\ and\
  \bibinfo {author} {\bibfnamefont{B.~F.}\ \bibnamefont{Whiting}},\ }%
  \bibfield{journal}{%
  \bibinfo {journal} {Phys.Rev.}\ }%
  \textbf{\bibinfo {volume} {D67}},\ \bibinfo {pages} {024025} (\bibinfo {year}
  {2003})%
  \bibAnnoteFile{NoStop}{Detweiler-Whiting:02}%
\bibitem{Dolan:11}%
  \BibitemOpen
  \bibfield{author}{%
  \bibinfo {author} {\bibfnamefont{S.~R.}\ \bibnamefont{Dolan}}\ and\ \bibinfo
  {author} {\bibfnamefont{L.}~\bibnamefont{Barack}},\ }%
  \bibfield{journal}{%
  \bibinfo {journal} {Phys.Rev.}\ }%
  \textbf{\bibinfo {volume} {D83}},\ \bibinfo {pages} {024019} (\bibinfo {year}
  {2011})%
  \bibAnnoteFile{NoStop}{Dolan:11}%
\bibitem{Poisson:05}%
  \BibitemOpen
  \bibfield{author}{%
  \bibinfo {author} {\bibfnamefont{E.}~\bibnamefont{Poisson}},\ }%
  \bibfield{journal}{%
  \bibinfo {journal} {Phys.Rev.Lett.}\ }%
  \textbf{\bibinfo {volume} {94}},\ \bibinfo {pages} {161103} (\bibinfo {year}
  {2005})%
  \bibAnnoteFile{NoStop}{Poisson:05}%
\bibitem{Dixon:74}%
  \BibitemOpen
  \bibfield{author}{%
  \bibinfo {author} {\bibfnamefont{W.~G.}\ \bibnamefont{Dixon}},\ }%
  \bibfield{journal}{%
  \bibinfo {journal} {Phil. Trans. Roy. Soc. Lond. A}\ }%
  \textbf{\bibinfo {volume} {277}},\ \bibinfo {pages} {59} (\bibinfo {year}
  {1974})%
  \bibAnnoteFile{NoStop}{Dixon:74}%
\bibitem{Steinhoff:10}%
  \BibitemOpen
  \bibfield{author}{%
  \bibinfo {author} {\bibfnamefont{J.}~\bibnamefont{Steinhoff}}\ and\ \bibinfo
  {author} {\bibfnamefont{D.}~\bibnamefont{Puetzfeld}},\ }%
  \bibfield{journal}{%
  \bibinfo {journal} {Phys.Rev.}\ }%
  \textbf{\bibinfo {volume} {D81}},\ \bibinfo {pages} {044019} (\bibinfo {year}
  {2010})%
  \bibAnnoteFile{NoStop}{Steinhoff:10}%
\end{thebibliography}%
\end{document}